\documentclass[apjl]{emulateapj}

\usepackage{amsmath}
\usepackage{apjfonts}
\usepackage{amssymb}
\usepackage{mathrsfs}
\usepackage{natbib}
\usepackage{color}
\def\citea#1{\citep{#1}}
\def\citea#1{}
\newcommand{\Msun}{\mbox{M$_\odot$}}

\begin{document}

\shorttitle{Distinguishing black holes and neutron stars}

\title{When can gravitational-wave observations distinguish between black holes and neutron stars?}

\author{
Mark Hannam\altaffilmark{1,4}, Duncan A. Brown\altaffilmark{2,4}, Stephen Fairhurst\altaffilmark{1,4}, 
Chris L. Fryer\altaffilmark{3,4}, and Ian W. Harry\altaffilmark{2,4} }

\altaffiltext{1}{School of Physics and Astronomy, Cardiff University, Cardiff, UK} 
\altaffiltext{2}{Department of Physics, Syracuse University, Syracuse NY} 
\altaffiltext{3}{Computational Computer Science Division, Los Alamos National Laboratory, Los Alamos, NM}
\altaffiltext{4}{Kavli Institute of Theoretical Physics, UC Santa Barbara, CA}

\begin{abstract}
Gravitational-wave observations of compact binaries have the potential to
uncover the distribution of masses and spins of black holes and
neutron stars in the universe.  The binary components' physical parameters can
be inferred from their effect on the phasing of the gravitational-wave signal,
but a partial degeneracy between the components' mass ratio and their spins 
limits our ability to measure the individual component masses.  At the
typical signal amplitudes expected by the Advanced Laser Interferometer
Gravitational-wave Observatory (signal-to-noise ratios between 10 and 20), we
show that it will in many cases be difficult to distinguish whether the
components are neutron stars or black holes.  We identify when the masses of
the binary components could be unambiguously measured outside the range of
current observations: a system with a chirp mass $\mathcal{M} \le 0.871\,$
M$_\odot$ would unambiguously contain the smallest-mass neutron star observed,
and a system with $\mathcal{M} \ge 2.786\,\Msun$ must contain a black hole.
However, additional information would be needed to distinguish between a
binary containing two $1.35$ M$_\odot$ neutron stars and an exotic
neutron-star--black-hole binary. We also identify those configurations that
could be unambiguously identified as black-hole binaries, and show how the
observation of an electromagnetic counterpart to a neutron-star--black-hole
binary could be used to constrain the black-hole spin.
\end{abstract}


\maketitle


\section{Introduction}
\label{sec:intro}

The first observing runs of the Advanced Laser Interferometer
Gravitational-wave Observatory (aLIGO) are expected in $\sim2015$, with Advanced
Virgo following on a similar schedule~\citep{Harry:2010zz,aVIRGO}.  The primary
source for these observatories is the coalescence of binaries containing black
holes and/or neutron stars, with predicted rates between a few and several
hundred per year at detector design
sensitivity~\citep{Abadie:2010cf,Dominik:2012vs}.  The components of these
binaries are formed in supernovae when the core of a massive star collapses
to a compact remnant, although the exact collapse mechanism remains
unknown.  Detailed knowledge of the mass distribution of black holes and
neutron stars will provide vital clues to their formation as well as explore
the equation of state (EOS) of nuclear matter at high densities.

Measuring the upper and lower limits of neutron-star masses allows us to
constrain the supernova engine and the nuclear physics of neutron-star
remnants~\citep{Lattimer:2010uk,Fryer12}.  For example, the collapse of
low-mass stars ($8 -12 \, \Msun$) is believed to quickly produce explosions with
very little mass accreted in a convective engine phase. The predicted mass of
the compact remnant will be less than the Chandrasekhar mass by an amount that
depends on the collapse model~\citep{Fryer99,Kitaura:2005bt,Dessart07}.  If
we can place an upper limit on the mass of a low
mass neutron star (in the $1.0 - 1.2 \, \Msun$ range), we can
distinguish between current models, effectively using these low-mass
systems to constrain the EOS and neutrino physics in
core-collapse.  At higher masses, \cite{Ozel:2010su} and \cite{Farr:2010tu}
have argued that there is a gap between $\sim 2- 4 \, \Msun$ where no compact
objects exist.  If true, the amount of material that falls back onto the newly
formed compact remnant must be small, arguing against engine mechanisms that
take a long time (more than 200\,ms) to develop~\citep{Belczynski:2011bn}.
However, this mass gap may simply be an artifact of poor mass resolution of
X-ray binaries and poor statistics~\citep{Kreidberg:2012ud}.  
Black-hole mass distributions will allow us
to explore the fall-back of material in a weak supernova explosion and
black-hole masses in solar metallicity environments will provide clues into stellar
mass loss.  With an accurate black-hole mass distribution, we can study these
open questions in stellar evolution.   

The binary's gravitational-wave phasing depends at leading order
on its chirp mass $\mathcal{M} = (m_1 m_2)^{3/5} (m_1 + m_2)^{-1/5}$, where
$m_{1}$ and $m_{2}$ are the binary's component masses~\citep{Peters:1963ux}; 
this quantity will be most accurately measured 
in a gravitational-wave detection. The mass ratio $\eta = m_1 m_2 / (m_1 +
m_2)^2$ enters through higher-order corrections and is less accurately
measured; see e.g.,~\cite{Blanchet:2002av}.  There is also a partial degeneracy
between the mass ratio and the angular momentum $\chi_{1,2} = J_{1,2} /
m_{1,2}^2$ of each compact object (the spin), which further
limits our ability to measure the binary's component masses. Heuristically,
this can be understood as follows: a binary with spins aligned with the
orbital angular momentum will inspiral more slowly than a non-spinning system.
Similarly, a binary of the same \textit{total mass} but with more extreme mass
ratio will inspiral more slowly.  However, a binary with the same
\textit{chirp mass} but with more extreme mass ratio will inspiral more
quickly.  The effect on the waveform of decreasing $\eta$ can be mimicked by
increasing the component spins. 

We investigate the accuracy with which the component masses
can be determined from gravitational-wave observations of binary-neutron-star
(BNS), neutron-star--black-hole (NSBH), and binary-black-hole (BBH) systems,
focussing on systems where the object's spins are aligned with the orbital angular 
momentum. 
Since the first signals detected by aLIGO are likely to have signal-to-noise
ratios (SNRs) close to the observable network threshold of
$\sim$12~\citep{Colaboration:2011np}, we focus on signals with SNRs 10--20, 
which will account for $\sim$80\% of observations.
For these SNRs, we find that the mass-ratio--spin degeneracy will prevent us
from accurately measuring component masses.
We identify the region of the mass parameter space for
which it will not be possible to determine whether the compact objects are
black holes or neutron stars using gravitational-wave observations alone, 
when we can conclusively measure compact-object masses outside the 
currently observed limits, and show how the observation of an electromagnetic 
counterpart to an NSBH could be used to constrain the black-hole spin.

\section{Parameter Estimation Method}
\label{s:method}

Fisher-matrix methods show that the binary's chirp mass is recovered well by
matched filtering, with accuracies of $\sim 0.01\%$ for typical BNS systems in
aLIGO~\citep{Finn:1992xs,Arun:2004hn}. If we assume that the neutron stars are
nonspinning 
$\eta$ can be
measured to an accuracy of $\sim 1.3\%$~\citep{Arun:2004hn}. Estimates of the
effect of the mass-ratio--spin degeneracy were first made by
\cite{Cutler:1994ys} and \cite{Poisson:1995ef} using the Fisher approach. The
degeneracy between the mass ratio and the total effective spin $\chi = (m_1
\chi_1 + m_2 \chi_2)/(m_1+m_2)$ degrades the ability to measure the mass ratio
and hence the component masses.  We go beyond these studies, using the method
introduced in \cite{Baird:2012cu} to equate a confidence interval with a
region where the match between the signal and model waveforms exceeds a given
threshold.  We use this method to investigate parameter degeneracies for a wide
range of binaries and interpret the expected measurement accuracy in the
context of the astrophysical questions discussed above.

We model the waveforms with the TaylorF2 inspiral approximant
\citep{Sathyaprakash:1991mt,Cutler:1994ys,Droz:1999qx} to leading order in 
amplitude and 3.5 post-Newtonian (PN) order in
phase~\citep{Blanchet:1995ez, Blanchet:2001ax, Blanchet:2004ek, Blanchet:2005a} 
with spin-orbit
terms to 3.5PN order and spin terms to 2PN order~\citep{Kidder:1992fr,Kidder:1995zr}.  
For systems with total
masses below $\sim8\,\Msun$, our results with TaylorF2 are consistent with those from
phenomenological BBH models that
include the merger and ringdown~\citep{Ajith:2009bn,Santamaria:2010yb},
calibrated against numerical-relativity waveforms with mass ratios up to
1:4~\citep{Hannam:2010ec}.  For the higher-mass BBH results in Sec.~\ref{s:bbh} we
use the full merger model. Throughout, we assume for simplicity that the
component spins are aligned with the orbital angular
momentum.  In this case, the binary's distance, orientation and sky location
affect only the overall amplitude of the waveform, and we do not consider them
here. 

For two waveforms $h_1$ and $h_2$ the match is given by
\begin{equation}
M = \max_{\Delta t, \Delta \phi} \frac{(h_1|h_2)}{\sqrt{(h_1|h_1)(h_2|h_2)}}
\end{equation}
where $(a|b)$ is the standard noise-weighted inner product 
\begin{equation}
(a|b) = 4 \, \mathrm{Re} \int_0^\infty \frac{\tilde{a}(f) \tilde{b}^\ast(f)}{S_n(f)}\,df.
\end{equation}
In all cases we use a noise sensitivity $S_n(f)$ corresponding to the 
zero-detuned high-power configuration of aLIGO~\citep{Shoemaker:aLIGO} 
with a 15~Hz low frequency cutoff.  We construct a
90\% confidence region for a signal in the $(m_{1}, m_{2},\chi)$  space, which
corresponds to the three-dimensional region where $M \ge 0.968$ $(0.992)$ for
an SNR of 10 (20)~\citep{Baird:2012cu}.  This method is
more accurate at low SNRs than the Fisher-matrix approach.

\section{Neutron-Star Binaries}
\label{s:bns}

\begin{figure}
\begin{center}
  \includegraphics[width=0.5\textwidth]{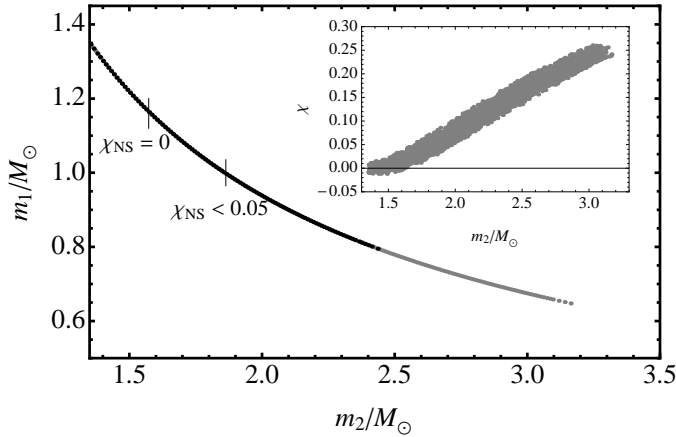}
  \caption{
  \label{fig:BNS} 
  The 90\% confidence region around a 1.35$\,\Msun$-1.35$\,\Msun$ 
  BNS system at SNR 10 (gray) and SNR 20 (black) in the 
  zero-detuned high-power configuration of aLIGO. 
  The vertical bars indicate where the SNR 10 confidence region would be
  truncated if we restrict to non spinning neutron stars, and if we restrict to neutron-star spins less 
  than 0.05. The inset shows the total effective spin $\chi = (m_1 \chi_1 + m_2 \chi_2)/(m_1+m_2)$ 
  of the waveforms, with respect to the mass of the larger body. 
  }
\end{center}
\end{figure}

Observed neutron-star masses currently lie between $1.0 \pm 0.10
\Msun$~\citep{Rawls:2011jw} and $1.97\pm 0.04 \Msun$~\citep{Demorest:2010bx}.
General relativity and causality place a strict upper limit on the maximum
neutron-star mass of $3.2 \Msun$~\citep{Rhoades:1974fn}; the actual maximum
mass is determined by the as yet unknown neutron-star EOS.  The
observed masses of stars in double neutron-star systems is narrower with a
peak at $1.35 \Msun$ and a width of $0.13 \Msun$ \citep{Kiziltan:2010ct}.  We
begin by considering a canonical BNS system with masses $m_1 = m_2 = 1.35
\Msun$ and no spins.  Fig.~\ref{fig:BNS} shows the regions of the mass plane that are
consistent with this source at 90\% confidence for SNRs 10 and 20.  The
component masses consistent with the signal lie roughly along a line of
constant chirp mass.  However, there is significant spread in the recovered
component masses due to the degeneracy between $\eta$ and $\chi_{1,2}$ in the
gravitational-wave phase evolution.  Typically, the degeneracy persists over a
range of $0.3$ in $\chi$.  

If we assume that the compact objects are non-spinning, the component masses
are recovered in the ranges $1.15$--$1.35 \Msun$ and $1.35$--$1.6 \Msun$. The
observed spins of double neutron stars are low, with the minimum observed
period of $22.70\, \mathrm{ms}$ for J0737-3039A \citep{Burgay:2003jj}, i.e.,
$\chi \sim 0.02$, where the neutron-star period can be related to the spin by 
approximately
\begin{equation}
\chi = \left( \frac{2\pi c I }{G m^2} \right)
\left( \frac{1}{T} \right) \approx 0.4 \left(\frac{1 \mathrm{ms}}{T} \right). 
\end{equation} 
If we constrain the neutron stars to have a spin $\chi \le
0.05$, the range of consistent component masses extends to $1.0$--$1.35
\Msun$ and $1.35$--$1.9 \Msun$.  Therefore at typical SNRs, we would be unable 
to distinguish the canonical $1.35 \Msun$ 
BNS from a more exotic BNS with $m_1 = 1.0\Msun, m_2 = 1.9\Msun$.

The fastest spinning pulsar (PSR J1748-2446ad) has a period of 1.4ms, or $\chi
\sim 0.3$~\citep{Hessels:2006ze}. Neutron stars are considered unlikely to have
a period less than 1~ms~\citep{Chakrabarty:2008gz}, although breakup frequencies
could be a factor of $\sim 2$ higher~\citep{Lo:2010bj}. If we allow for
larger component spins, the region consistent with the canonical BNS system
extends to $0.65$--$1.35 \Msun$ and $1.4$--$3.1\Msun$.  The maximum
total spin of binaries within the 90\% confidence region at SNR 10 is
$\sim$0.3; see Fig.~\ref{fig:BNS} inset.  Based on
gravitational wave observations alone, a binary of two $1.35 \Msun$ neutron
stars could not be definitively distinguished from a low-mass neutron star and
a black hole or neutron star in the mass gap, even at SNR 20.   If the larger object is a neutron
star its spin would be unusually large, but not impossible. To constrain the
larger object's mass below 2\,$\Msun$ without assumptions on the spins would
require a signal with an SNR of $\gtrsim$40, expected for $\sim 2\%$ of observations.

\begin{figure}
\begin{center}
  \includegraphics[width=0.5\textwidth]{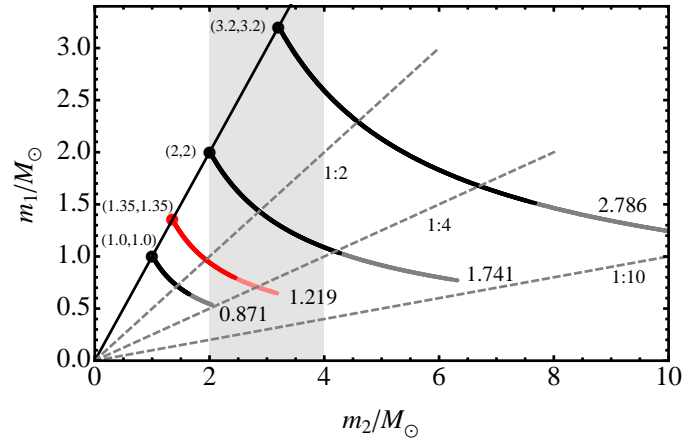}
  \caption{
  \label{fig:CMlines} 
  The 90\% confidence regions for a number of different non-spinning 
  compact-binary configurations; see text for interpretation. The number at the 
  end of each confidence region is the chirp mass of the binary. The gray 
  shaded region indicates the current observational mass gap.
  }
\end{center}
\end{figure}

In Fig.~\ref{fig:CMlines}, we consider a family of equal mass, non-spinning
BNS systems with component masses at the lowest ($1.0 \, \Msun$) and highest
($2.0 \, \Msun$) observed neutron-star masses, and at the highest theoretical
neutron-star mass ($3.2 \, \Msun$). The $1.35 \, \Msun$ binary is shown for
reference, and the observed mass gap is indicated by the shaded region.
Again, the 90\% confidence regions follow lines of approximately constant
chirp mass.  We can see that for an exceptionally low or high mass binary, we
will be able to identify \textit{at least one} of the components as
extraordinary.  The observation of a BNS with a chirp mass $\mathcal{M} \le
0.871 \, \Msun$ would yield the unambiguous detection of a compact object with mass less
than $1.0 \, \Msun$.  The less
massive component could have a mass in the range $0.5$--$1.0 \, \Msun$, but any
mass in that range would challenge our current understanding of neutron stars
and their formation in supernovae. 
A chirp mass $\mathcal{M} < 1.045$ would yield an unambiguous detection of a 
$<1.2\,\Msun$ neutron star.  This
constraint would rule out many modern calculations of stellar 
collapse~\citep{Kitaura:2005bt,Dessart07} and provide a strong validation test for 
future calculations.

Similarly, the observation of a system
with $\mathcal{M} \ge 1.741 \, \Msun$ would indicate the detection of a neutron star of
mass $\ge 2.0 \, \Msun$ \textit{provided we can unambiguously identify both
components as neutron stars}.  However, the observation of a BNS system with
$m_1 = m_2 = 2.0 \, \Msun$ is consistent with a binary containing a neutron star
and a black hole in the mass gap or an exotic, low-mass neutron star and a
black hole.  Although any configuration along this line is of interest, we
cannot strongly constrain the component masses with gravitational-wave
observations at low SNR. In particular, even if we assume that neutron stars are
nonspinning and that the minimum BH mass is 3\,$\Msun$, 
we cannot rule out that the possibility that it is a NSBH system: our assumption would 
only remove the portion of the confidence interval 
with $2.4\,\Msun \leq m_2 \leq 3.0\,\Msun$, and the observation would be of \emph{either} a
BNS system with well-constrained masses, \emph{or} an NSBH.

Finally, we consider a binary with component masses at the upper end of
the neutron-star mass limit, above which we expect the components to be black
holes.  In this case, we could conclusively say that one of the components
must be a black hole, although the degeneracy limits our ability to draw
strong conclusions on the component masses. The binary could either be a BBH
system with component masses in the mass gap, or an NSBH system with masses
consistent with previously observed compact objects. 

Note that in all the cases we have considered, the equal-mass line provides
a hard upper (lower) limit on $m_1$ ($m_2$).

Fig.~\ref{fig:CMlines} shows that it will be difficult to explore the mass
gap with single gravitational-wave observations at low SNRs. To constrain one
of the masses to lie within the mass gap, the system would need to be observed
with an SNR of greater than 30 ($\sim 5\%$ of observations); in the case of the 
$m_1 = m_2 = 2.0 \, \Msun$
binary, the mass on one of the objects would then be constrained between
$2\,\Msun$ and $3.5\,\Msun$, placing it directly within the mass gap. 

\section{Binary Black Holes}
\label{s:bbh}

Black hole spins can vary between $0$ and $1$, with observations of X-ray
binaries supporting the full range of values~\citep{Zhang:1997dy}.  As before, 
the mass-ratio--spin degeneracy precludes precise measurement of the component
masses.  However, if we observed a BBH with $m_1 = m_2 = 36\,$M$_\odot$, 
corresponding to a binary with both components above $35\,\Msun$---the largest mass 
observed for a stellar-mass black hole in an X-ray binary~\citep{Silverman:2008ss}--- 
we \emph{could} conclude that one of the black holes has a
mass of $36\,\Msun$ or higher, providing the first observational evidence for
a stellar-mass black hole above $35\,\Msun$. The same is true for any binary
with the same or larger chirp mass, $\mathcal{M} \ge 31.34\,\Msun$.

\begin{figure}
\begin{center}
  \includegraphics[width=0.5\textwidth]{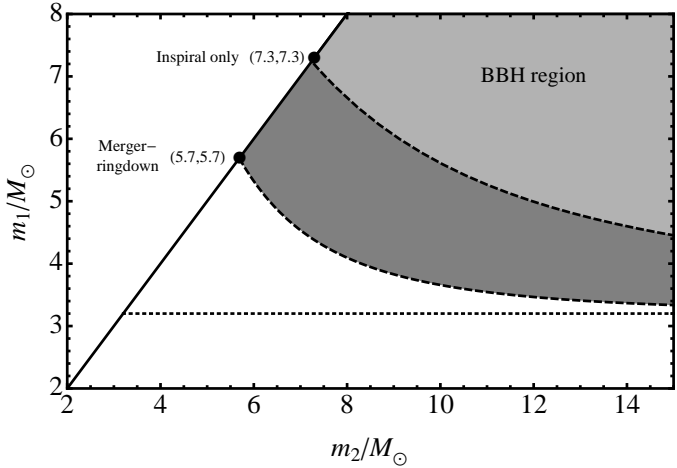}
  \caption{
  \label{fig:BHmassrange} 
  The shaded region indicates binaries
  would be unambiguously identified as BBH systems, assuming a maximum neutron
  star mass of $3.2\,\Msun$ (dotted line). The light gray region represents the most
  conservative estimate of this region, using an inspiral-only waveform model, while
  the additional darker region uses a full merger-ringdown model. 
    }
\end{center}
\end{figure}

A BBH will only be unambiguously identified if all binaries within the
90\% confidence interval for the measured component masses are also BBH systems.  The
shaded region in Fig.~\ref{fig:BHmassrange} shows the part of parameter space
in which BBH systems could be identified as such at SNR 10, assuming a
maximum neutron-star mass of $3.2\,\Msun$. The upper curve shows the boundary
of this region computed using the inspiral only (TaylorF2) model. The lower
curve is computed using a phenomenological inspiral-merger-ringdown
model~\citep{Santamaria:2010yb}. When the additional information from merger
and ringdown is included, a larger region of the parameter space can be
identified as a BBH system, illustrating the impact of merger-ringdown in these
systems. 

\section{Electromagnetic counterparts}
\label{s:nsbh}

\begin{figure}
\begin{center}
  \includegraphics[width=0.5\textwidth]{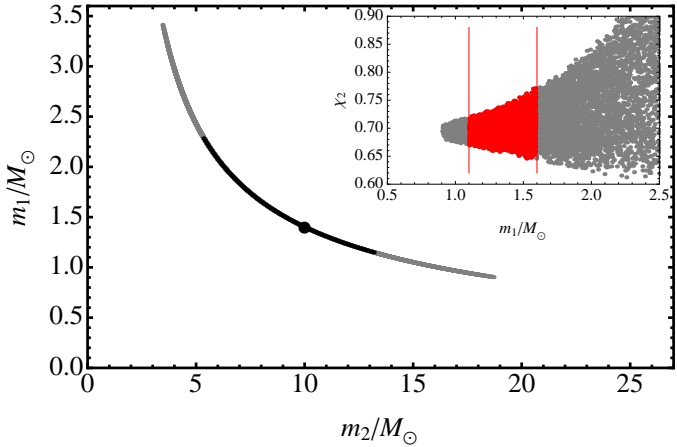}
  \caption{
  \label{fig:BHspin} 
  The 90\% confidence regions at SNRs 10 and 20 for an NSBH system with 
  masses $m_1 = 1.4 \Msun$ and $m_2 = 10 \Msun$, and BH spin 
  $\chi_2 = +0.7$. A GRB observation would allow an estimate of the 
  BH mass of $m_2 = 11\pm3 \Msun$, and $\chi_2 = 0.7\pm0.05$ (see
  text).
    }
\end{center}
\end{figure}

Fig.~\ref{fig:BHspin} shows the 90\% confidence regions for an NSBH, where the
NS has mass 1.4~$\Msun$ and no spin, and the BH has $m_{2} = 10 \Msun$ with
spin $\chi_{2} = 0.7$ aligned with the orbital angular momentum.  The
uncertainty in both masses is large: this could be an NSBH system
containing a BH between $6 \, \Msun$ and $26 \, \Msun$, or an equal-mass BBH system,
with masses $m_1 = m_2 = 3.5 \, \Msun$.  The possible mass ratio of this system
extends from 1:1 through to 1:25. Even at SNR 20 the mass ratio ranges from
1:2.4 up to 1:11.  We note that the waveform models used here only include the
waveform's dominant harmonic and do not include merger and 
ringdown (the phenomenological model is calibrated only up to mass ratio 1:4). 
The confidence region may significantly change if
higher-order or merger-ringdown effects are included, emphasizing the need for
further waveform development in interpreting gravitational-wave observations
as well as for detection.

The observation of a gamma-ray burst or other electromagnetic signal would
unambiguously identify one of the components as a NS and break the 
mass-ratio--spin degeneracy.  If we accept that the mass of a neutron star in a binary is
$1.35\pm0.25 \Msun$~\citep{Kiziltan:2010ct}, then the BH mass is restricted to
$m_2 = 11\pm3 \Msun$. The further assumption that the neutron-star spin is
below 0.4 restricts the possible values of the black-hole spin, allowing an
estimate of $\chi_2 = 0.7\pm0.05$. This is shown in Fig.~\ref{fig:BHspin}
(inset), which shows the 90\% confidence region restricted to the
$\chi_2$-$m_1$ plane, and the region for which the neutron-star mass is
between $1.1 \Msun$ and $1.6 \Msun$. However, due to beaming only a fraction of 
gravitational-wave observations of BNS and NSBH systems are likely to be accompanied by GRB 
observations~\citep{Briggs:2012ce}.

\section{Conclusions}
\label{s:conclusion}

Although  gravitational-wave observations will accurately measure the chirp
mass $\mathcal{M}$ of binary mergers, a mass-ratio--spin degeneracy prevents the component
masses and spins being measured accurately at low SNRs. In many cases
it will be difficult to determine whether the components of the
binary are neutron stars or black holes. However, we have illustrated several
cases where significant results can be inferred from gravitational-wave
observations.

We illustrate several situations where the binary must contain (at least one)
compact object that is more or less massive than anything observed
to date; $\mathcal{M} < 0.871\,\Msun$ indicates a neutron star smaller than
1\,$\Msun$, $\mathcal{M} > 1.741\,\Msun$ a neutron star larger than 2\,$\Msun$
(if we can independently verify that it is a BNS system), and $\mathcal{M} >
31.34\,\Msun$ a black hole larger than 35\,$\Msun$.  Observations
at SNRs higher than 30 will be required to clarify the existence of the mass
gap.  The observation of an electromagnetic counterpart will, in certain
situations, allow us to identify the system as an NSBH and, if we know the
distribution of neutron-star masses, measure the black-hole mass and spin with
high accuracy.  

Our results are qualitatively robust, but the true confidence intervals may vary; they 
were generated by calculating waveform mismatches~\citep{Baird:2012cu}, 
rather than full parameter-estimation 
methods~\citep{Sluys:2008a, Sluys:2008b, Veitch:2010, Feroz:2009}. In some
cases inclusion of merger-ringdown, higher-order post-Newtonian corrections to 
spin effects, particularly at high mass ratios~\citep{Arun:2008kb}, and tidal 
effects will slightly alter the size of the error regions.  
The issues only
serve to highlight the urgency of improved waveform models.
The inclusion of precession and higher harmonics will modify our
results, but we do not expect these to significantly break the degeneracy
we discuss~\citep{Baird:2012cu}.

Our results highlight the significance of the mass-ratio--spin degeneracy in
gravitational-wave observations of compact binaries. This needs to be explored
further using more complete parameter-estimation methods, more accurate
waveform models, and extended to include an exhaustive study of the
parameter space. Results from a population of signals, rather than
individual observations, should also be investigated. Understanding
the uncertainties in masses and spin measurements will be
essential to interpreting gravitational-wave observations. 


\acknowledgements 

We thank Harald Pfeiffer, Mark Scheel and Patrick Sutton for helpful discussions. 
DB is supported by NSF award PHY-0847611 and an RCSA Cottrell Scholar award. 
SF is supported by the Royal Society. 
MH is supported by STFC grants ST/H008438/1 and ST/I001085/1. 
IH is supported by NSF grants PHY-0847611 and PHY-1205835.
This work was funded in part under the auspices of the U.S. Dept. of Energy, and supported 
by its contract W-7405-ENG-36 to Los Alamos National Laboratory.   
We thank the Kavli Institute for Theoretical Physics at
UC-Santa Barbara, supported in part by NSF grant PHY11-25915, where
this work was conceived. 



\end{document}